\documentclass[11pt,english]{article}
\usepackage{times}
\usepackage[T1]{fontenc}
\usepackage[latin1]{inputenc}
\usepackage{geometry}
\usepackage{graphicx}

\makeatletter

\date{ISMRM  2004, Poster 2323, Thursday, May 20}

\usepackage{epsfig}
\usepackage{color}
\definecolor{dgray}{rgb}{0.3,0.3,0.3}%   rgb color model
\definecolor{lgray}{rgb}{0.5,0.5,0.5}%   rgb color model

\usepackage{babel}
\makeatother
\begin{document}

\title{Effects of $T_{2}$ Relaxation and Diffusion on Longitudinal Magnetization
State and Signal Build for HOMOGENIZED Cross Peaks}

\author{{\Large Curtis A. Corum$^{\textrm{1}}$, Arthur F. Gmitro$^{\textrm{1,2}}$}}

\maketitle
{\large $^{\textrm{1}}$Optical Sciences Center~(corum@email.arizona.edu), }{\large \par}

{\large $^{\textrm{2}}$Department of Radiology, University of Arizona,
Tucson, AZ, USA}{\large \par}

\begin{abstract}
An analytical expression has been developed to describe the effects
of $T_{2}$ relaxation and diffusing spatially modulated longitudinal
spins during the signal build period of an HOMOGENIZED cross peak.
Diffusion of the longitudinal spins results in a lengthening of the
effective dipolar demagnetization time, delaying the re-phasing of
coupled anti-phase states in the quantum picture. In the classical
picture the unwinding rate of spatially twisted magnetization is no
longer constant, but decays exponentially with time. The expression
is experimentally verified for the HOMOGENIZED spectrum of 100mM TSP
in $H_{2}O$ at 4.7T.
\end{abstract}

\section*{Introduction}

HOMOGENIZED\cite{VLW96} and its variants\cite{CHC+04,FB04} and the
recently proposed IDEAL\cite{ZCC+03} sequences have great potential
for in-vivo spectroscopy\cite{FPH03}. Diffusion weighting in HOMOGENIZED
is present both to give intentional diffusion weighting and as a side
effect of the various gradients present. Stekjsal-Tanner (ST) diffusion
weighting\cite{ST65} during the $t1$ (mix) and $t2$ (build) periods
of the sequence can be used to suppress radiation dampening. Enhanced
diffusion weighting\cite{JZ01,ZCJZ01,ZCGL01} is obtained during $t1$.
There is an additional $t2$ dependent diffusion weighting possible,
due to the iZQC gradient $G_{zq}$ and $\beta$ pulse combination.
The weighting results from diffusing modulated longitudinal magnetization,
and does not behave as ST diffusion weighting. Kennedy et al.\cite{KRC+03}
have shown recently that this diffusion weighting has the novel property
of being insensitive to object motion.

\begin{figure}
\includegraphics[%
  width=1.0\columnwidth,
  keepaspectratio]{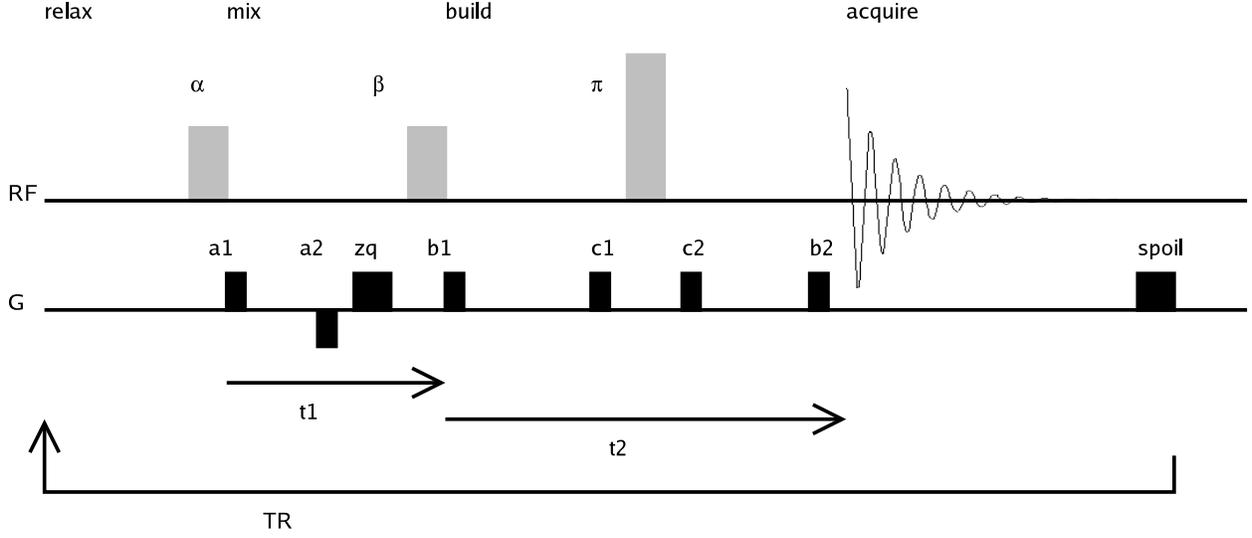}

\caption{\label{cap:Pulse-Sequence}Pulse Sequence. $\alpha$, $\beta$, and
$\pi$ RF pulses are all the same phase}
\end{figure}

We have concentrated our efforts on the 2d HOMOGENIZED sequence shown
in figure \ref{cap:Pulse-Sequence}. 

The sequence consists of three RF pulses, $\alpha$ for excitation,
\textbf{$\beta$} to convert helical transverse magnetization to $M_{z}$
modulation, and $\pi$ to form a spin echo. The $G_{a1}$, $G_{a2}$
gradient pair yields ST diffusion weighting and radiation damping
suppression during $t1$. The $G_{b1}$, $G_{b2}$ gradient pair accomplishes
the same during $t2$. The $G_{c1}$, $G_{c2}$ gradient pairs crush
transverse magnetization created at $\pi$ due to pulse imperfections
and $B_{1}$ inhomogeneity, as well as introduce some additional diffusion
weighting. The $G_{zq}$ gradient selects intermolecular zero quantum
coherences (iZQCs) in the quantum picture. In the classical picture
the $G_{zq}$ gradient in combination with $\beta$ creates spatially
modulated longitudinal magnetization whose magnetic field causes unwinding
(and eventually rewinding) of helically twisted transverse magnetization\cite{CG04c}.

\begin{figure}
\includegraphics[%
  width=1.0\columnwidth,
  keepaspectratio]{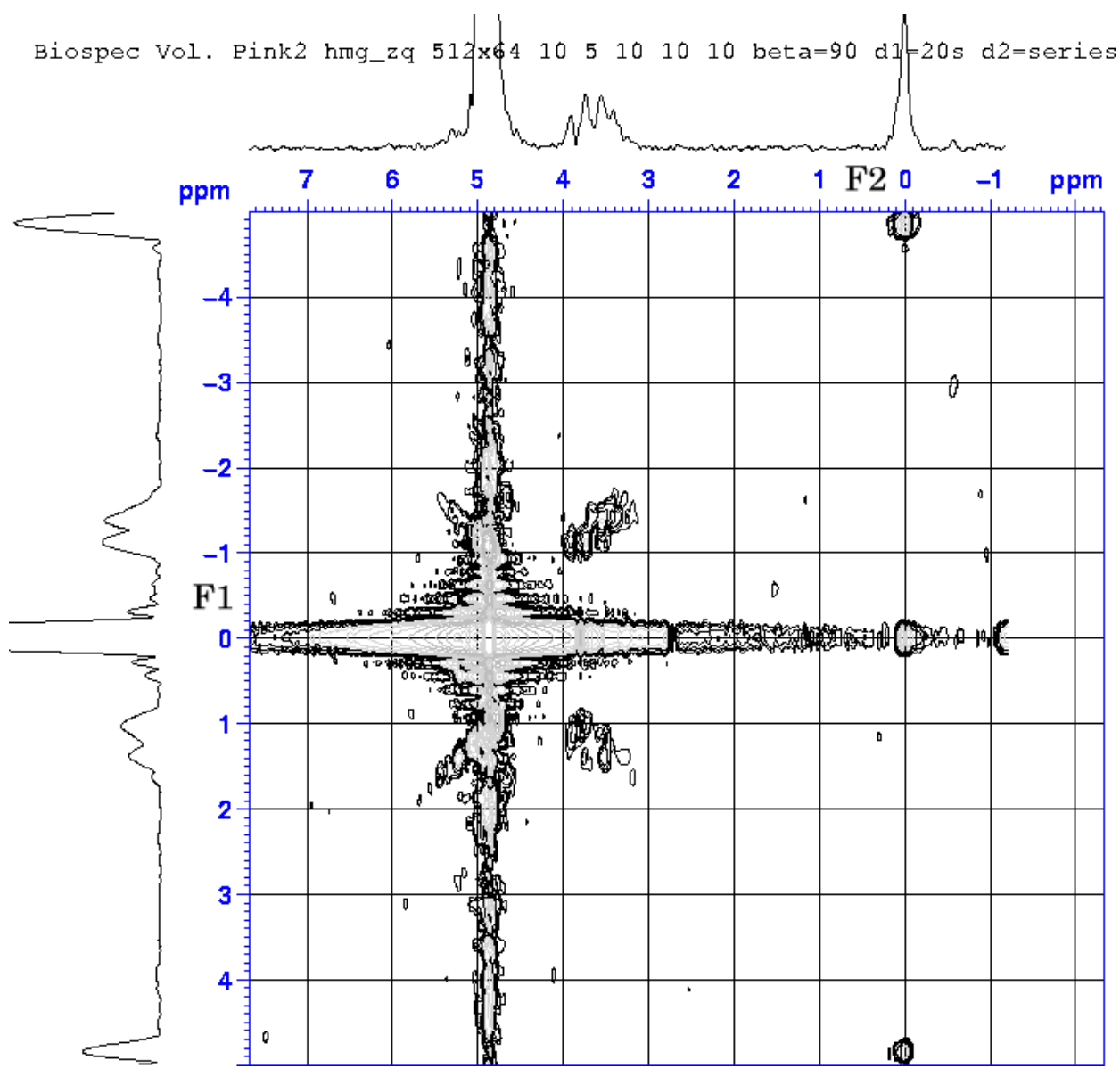}

\caption{\label{cap:spectrum}Representative low resolution 2d HOMOGENIZED
spectrum. TSP is referenced to -4.7ppm on F1 axis and 0.0ppm on F2
Axis. Projections are restricted to {[}0, 4{]} ppm F2 and {[}-5, -1{]}
ppm F1.}
\end{figure}

A 2d HOMOGENIZED experiment (see figure \ref{cap:spectrum}) is carried
out by incrementing $t1$ over multiple acquisitions. In a two (or
multiple) component system where only one component is present in
high concentration (the solvent $S$), cross peaks with the solute
component of interest $I$ will be formed at $(F1,\, F2)=(I-S,\, I)$
the $"p"$ type iZQC and $(S-I,\, I)$ the $"n"$ type iZQC. Axial
iZQC peaks are formed at $(0,\, I)$ and $(0,\, S)$. The equation
for peak amplitudes, neglecting radiation damping, $T_{2}$ relaxation,
and diffusion has been described in the literature\cite{alw98b}.

\section*{Theory}

For the first time, to the authors' knowledge, an analytical expression,
equation (\ref{eq:crosspeak}), for the cross peak amplitude in the
presence of diffusion and $T_{2}$ relaxation has been developed.
Some preliminary definitions precede the expression, notation follows
Ahn et al.\cite{alw98b}\begin{equation}
q_{zq}\equiv\frac{\gamma\, G_{zq}\delta_{zq}}{2\pi}\label{eq:q}\end{equation}
is the spatial frequency of periodic longitudinal magnetization $M_{z}$
formed by $G_{zq}$ of duration $\delta_{zq}$ and RF pulse $\beta$.

\begin{equation}
\tau_{Seff}\equiv\tau_{S}e^{(b_{a}+b_{zq})\, D_{S}}e^{\frac{t1}{T_{2}^{S}}}\label{eq:tau}\end{equation}
$\tau_{Seff}$ (\ref{eq:tau}) has been defined to take account of
$T_{2}$ and diffusion losses (ST b-values, $b_{a}$ and $b_{zq}$)
incurred during $t1$ before $\beta$ beta forms $M_{z}$. $\tau_{S}$
is the dipolar demagnetization time for spin S as per reference \cite{alw98b}. 

\begin{equation}
F(t2)\equiv\frac{1-e^{-t2(2\pi\, q_{zq})^{2}D_{S}}}{\tau_{Seff}(2\pi\, q_{zq})^{2}D_{S}}\label{eq:F}\end{equation}
can be thought of as an exponentially slowing ''winding'' parameter,
instead of the linear (in $t2)$ winding parameter $\frac{t2}{\tau_{S}}e^{-\frac{t1}{T_{2}^{S}}}$
when diffusion is negligible. The new analytical expression for the
signal amplitude in the presence of diffusion and $T_{2}$ decay is

\begin{equation}
M_{p}=M_{0}^{I}e^{-(b_{a}+b_{zq}+b_{c}+b_{b})\, D_{I}}e^{-\frac{(t1+t2)}{T_{2}^{I}}}[\frac{cos(\beta)+1}{2}]\, J_{1}[sin(\beta)\frac{2}{3}F(t2)],\label{eq:crosspeak}\end{equation}
where $M_{p}$ is the p-type cross peak amplitude. $b_{a}$ , $b_{b}$,
and $b_{c}$ are the ST b-values due to the $G_{a}$, $G_{b}$, $G_{c}$
gradient pairs, respectively. $G_{zq}$ also introduces some ST diffusion
weighting $b_{zq}$ in the short delay before $\beta$.

\begin{figure}
\includegraphics[%
  width=1.0\columnwidth,
  keepaspectratio]{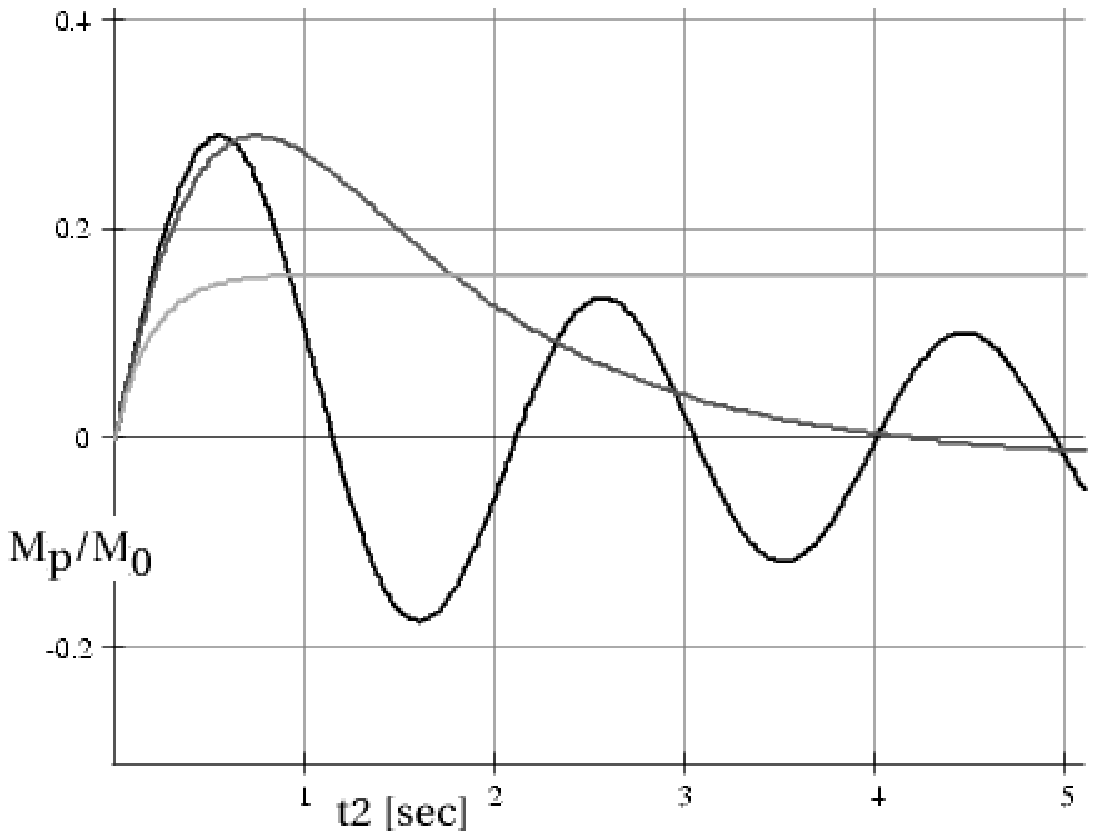}

\caption{\label{cap:Plot--theoretical}Plot of theoretical cross peak amplitude
$M_{p}$ vs. $t2$, for the case of negligible $T_{2}$ decay. $\beta=90^{\circ}$
and $\tau_{S}=200ms$. Three situations are shown:}

Black - negligible diffusion

\textcolor{dgray}{Dark Gray} - diffusion of $M_{z}$ has delayed
the maximum and stretched the zero crossings to longer times.

\textcolor{lgray}{Light Gray} - $M_{z}$ modulation has completely
diffused away before the maximum can be obtained.
\end{figure}

The effect of $F(t2)$ is to stretch the time axis when diffusion
weighting is significant. Equation (\ref{eq:crosspeak}) is valid
as long as $S$ and $I$ are separated by $1/\tau_{S}$ in frequency,
so that only longitudinal $S$ magnetization contributes to signal
build. Steady state values $(TR<5\, T_{1}^{S}or\, T_{1}^{I})$ may
be used for $\tau_{S}$ and $M_{0}$, as long as diffusion has eliminated
residual spatial modulation of longitudinal magnetization\cite{CG04}.
As long as the $a$ and $b$ gradient areas are chosen correctly,
radiation dampening is not significant. Three theoretical situations
are shown in figure \ref{cap:Plot--theoretical}. A similar expression
has been found for the n-type crosspeak amplitude $M_{n}$.

\section*{Experimental Results}

\begin{figure}
\includegraphics[%
  width=1.0\columnwidth,
  keepaspectratio]{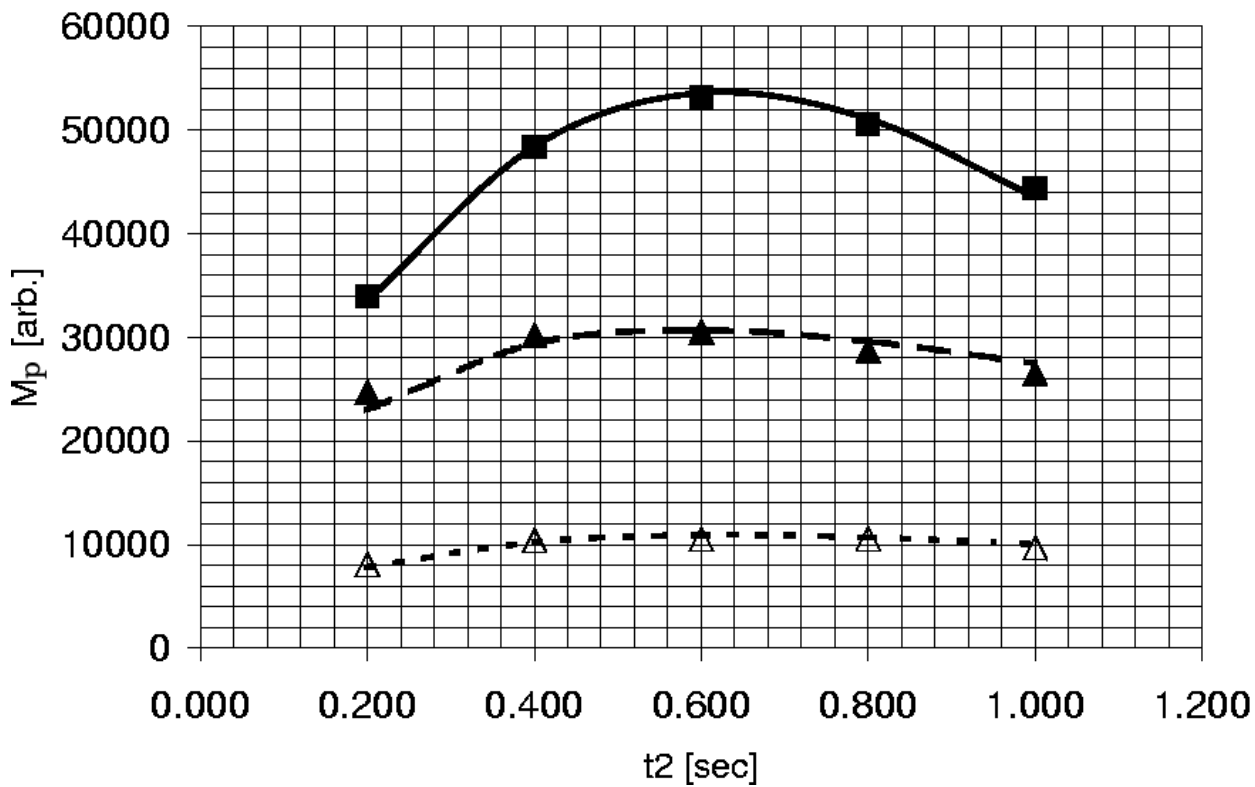}

\caption{\label{cap:Data-points}Data points and theoretical curve of p type
TSP peak for three cases. Y axis arbitrary units.Data points and theoretical
curve of p type TSP peak for three cases. Y axis arbitrary units.}

$\alpha=\beta=90^{\circ}$, $\delta_{a}=\delta_{b}=\delta_{c}=1ms$,
$\delta_{spoil}=5ms$

$G_{a}=G_{b}=G_{c}=G_{spoil}=20\frac{mT}{m}$, $\delta_{zq}=3ms$

Upper - $TR=20s,\, G_{zq}=10\frac{mT}{m}$

Middle - $TR=20s,\, G_{zq}=40\frac{mT}{m}$

Lower - $TR=2s,\, G_{zq}=40\frac{mT}{m}$
\end{figure}

A series of low resolution (512x64) HOMOGENIZED spectra were obtained
with various strengths of $G_{zq}$ (see figure \ref{cap:spectrum}).
The solvent (S) is water at room temperature, the solute of interest
(I) was TSP at 100mM concentration. Glucose was also present in solution.
Field strength is 4.7T yielding nominal $\tau_{S}=200ms$. A best
fit, adjusting $M_{0}^{I}$and $\tau_{S}$ to account for pulse imperfections
and $B_{1}$ inhomogeneity, was obtained for the top curve, and kept
the same for the other curves. Relaxation rates were measured in separate
inversion recovery and spin echo experiments with $T_{1}^{S}=2.57s$,
$T_{2}^{S}=140ms$ and $T_{2}^{I}=1.62s$. Effects such as $B_{1}$
inhomogeneity and RF pulse error contribute to lengthen $\tau_{Seff}$
(reduce available S magnetization). Comparison of the predicted cross
peak amplitude with experiment is shown in figure \ref{cap:Data-points}.
An analysis of axial peaks and $T_{1}$ effects is in progress\cite{CG04,CG04d}.

\section*{Acknowledgements}

This work and preparation leading to it was carried out under the
support of the Flinn Foundation, a State of Arizona Prop. 301 Imaging
Fellowship, and NIH 5R24CA083148-05.

\section*{Notes}

There is an error in the conference abstract on CD for the equation
for $M_{p}$. $cos(\beta)$ appears in the abstract where $\frac{cos(\beta)+1}{2}$
is the correct term as in equation (\ref{eq:crosspeak}). 

An extensive DDF/iMQC bibliography can be found at \cite{CC02}.

{\small \bibliographystyle{plain}
\bibliography{/home/corum/references}
}
\end{document}